\documentclass[preprints,article,accept,pdftex,moreauthors]{Definitions/mdpi} 
\firstpage{1} 
\pubvolume{1}
\issuenum{1}
\articlenumber{0}
\pubyear{2022}
\copyrightyear{2022}
\externaleditor{Academic Editor:}
\datereceived{18 May 2022} 
\dateaccepted{21 June 2022} 
\datepublished{} 
\datecorrected{} 
\dateretracted{} 
\hreflink{https://doi.org/} 
\pdfoutput=1

\usepackage[export]{adjustbox}

\Title{Shore Shadow Effect in Baikal}
\TitleCitation{Shore Shadow Effect in Baikal}

\Author{{Apoorva Bhatt} \orcidA{}, Pawe{\l} Malecki \orcidB{} and Dariusz G\'ora *\orcidC{}}

\AuthorNames{Apoorva Bhatt, Pawe{\l} Malecki and Dariusz Gora}

\AuthorCitation{{Bhatt, A.}; Malecki, P.; G\'ora, D.}
\address[1]{Institute of Nuclear Physics Polish Academy of Sciences, 31-342 Krakow, Poland; {apoorva.bhatt@ifj.edu.pl}, {pawel.malecki@ifj.edu.pl}}

\corres{\hangafter=1 \hangindent=1.05em \hspace{-0.82em}Correspondence: dariusz.gora@ifj.edu.pl}

\abstract{The measurement of the individual charged particles especially muons in an extended air shower (EAS) resulting from primary cosmic rays provides important distinguishing parameters to identify the chemical composition of the cosmic primary particles. For Neutrino Telescope experiments like Baikal-GVD, the estimation of underwater muon flux is of importance to study atmospheric muons. In this paper, a GEANT4-based simulation is presented to estimate the atmospheric muon flux underwater taking Baikal-GVD as an example. The location of the Baikal-GVD experiment at Lake Baikal provides a unique opportunity to study the passage of muons through its northern shore and the water. The muons arriving from the north direction will lose more energy as compared to those arriving from the south. An approximation for the northern shore is also simulated in the GEANT4 geometry and the results of the simulation are compared with the measurements from the NT-96 detector. The results of the simulations are consistent with the shore shadow observed in the measurements in the NT-96. This approach can also be used to propagate the muons from generators like CORSIKA through long distances in matter like water, ice, earth, etc for simulations in such experiments.}
\keyword{cosmic ray muons; simulation; neutrino telescope} 

\begin{document}
	
	
	\section{Introduction}
	
	Cosmic rays are high-energy particles that originate in outer space. The main sources of the cosmic rays which strike the earth's atmosphere from all directions are the Sun, objects in our Galaxy, and objects beyond our Galaxy, depending on the energy of the primary cosmic ray. The cosmic rays are mostly nuclei of atoms such as H, He, and heavier elements. They also include $e^-$, $e^+$, and other subatomic particles. The energy of cosmic rays varies from a few MeVs {(10$^6$ eVs)} to TeVs {(10$^{12}$ eVs)} and beyond. The primary cosmic rays interact with the atoms and nuclei in the atmosphere producing large cascades of secondary particles mainly hadrons, known as the Extensive Air Showers (EAS). The hadrons undergo strong interactions with the atmospheric nuclei such as nitrogen and oxygen and produce hadron showers containing mainly $\pi$'s and $K$'s. Out of the secondary particles, $\pi$'s are the most abundant due to their lower mass. If the secondary particles have sufficient energy they initiate new interactions. The unstable particles like $\pi$s and $K$s decay through the channels,
	\begin{equation}
		\pi^+\left(\pi^-\right) \rightarrow \mu^+\left(\mu^-\right) + \nu_\mu\left(\bar\nu_\mu\right),
	\end{equation}
	\begin{equation}
		K^+\left(K^-\right) \rightarrow \mu^+\left(\mu^-\right) + \nu_\mu\left(\bar\nu_\mu\right),
	\end{equation}
	{Since} muons are minimum ionizing particles they can penetrate large amounts of matter whereas weakly interacting neutrinos pass through matter with hardly any interaction. The muons are the most abundant charged particles at sea level. The muons lose approximately 2 GeV {(10$^9$ eV)} to ionization energy loss before reaching the ground. For very high energies of the muon, the energy loss is also dominated by the radiative processes resulting in much higher energy losses. The mean energy of the muons at the sea level is about $\sim$4 GeV. The integral intensity of vertical muons above 1 GeV at the sea level is $\approx$70 m$^{-2}$ s$^{-1}$ sr$^{-1}$. These muons have energies ranging from MeVs to TeVs. The low-energy muons decay into electrons and neutrinos.
	\begin{equation}
		\mu^+\left(\mu^-\right) \rightarrow e^+\left(e^-\right) + \nu_e\left(\bar\nu_e\right) + \bar\nu_\mu\left(\nu_\mu\right),
	\end{equation}
	
	The muons in an EAS are sensitive to the primary composition of cosmic rays and to the hadronic interaction properties. The muons are produced essentially from decaying charged mesons and can be a fundamental tool to map the high energy hadronic interaction, as the individual particle signature is not washed out as in the electromagnetic component. In a typical EAS, muons represent about 10 \% of all the charged particles. For negligible muon decay rate ({$E_\mu$} $>$ 100 $\left[\mathrm{GeV}\right]/\mathrm{cos}\theta$), the overall muon number spectrum can be expressed by an approximate extrapolation formula \cite{GaisserBook}
	\begin{equation}
		\frac{dN_\mu}{dE_\mu d\Omega} = \frac{0.14E_\mu^{-2.7}}{\mathrm{cm}^2\cdot \mathrm{sr}\cdot \mathrm{GeV}}\left(\frac{1}{1+\frac{1.1E_\mu cos\theta}{115 \left[\mathrm{GeV}\right]}} +\frac{1}{1+\frac{0.054E_\mu cos\theta}{850 \left[\mathrm{GeV}\right]}}\right),
	\end{equation}
	{Here}, the two terms give the contribution of pions and charged kaons, respectively. The contribution from the charm and heavier flavors are negligible except at very high energies and is neglected in this formula. A large fraction of the muons reach the ground due to a low interaction cross-section and a long decay time, and they have a wide lateral distribution. The measurement of the individual charged particles especially muons in an EAS would provide important distinguishing parameters to identify the chemical composition of cosmic primary particles.

	The measurements of muon multiplicity by various experiments on the surface, underground and underwater can be used to improve the parameters for the hadronic models at higher energies, also to study the cosmic ray spectral index and the composition of primary cosmic rays \cite{2020Surya}. The measurements of atmospheric muon flux are also used to improve the predictions of the atmospheric neutrino flux which is an important input to studying neutrino oscillations in atmospheric neutrino experiments. In the study of azimuthal dependence of muon flux discussed by Pethuraj et al. \cite{2020Pethuraj}, during the simulation of atmospheric muons, the detector along with the experimental hall building as well as buildings surrounding the experimental hall were simulated to account for all the materials the muon is traversing before reaching the detector. In the low energy region, these can give significant uncertainties due to muon energy losses.
	
	For the experiment at Baikal lake, the muon angular distributions, as well as the depth dependence of the vertical muon flux, were studied in the data taken with NT-36 to NT-96 detector \cite{2000BaikalNT}. In the work done by Balkanov et al. \cite{2000BaikalNT}, the experimental results of the shore shadow effect were presented as the ratio of zenith angle distribution of muons from the north and south. A sharp deficit in the number of muons arriving from the north was observed in the zenith angles larger than 70$^\circ$.
	
\begin{figure}[H]
		\includegraphics[width=0.8\linewidth]{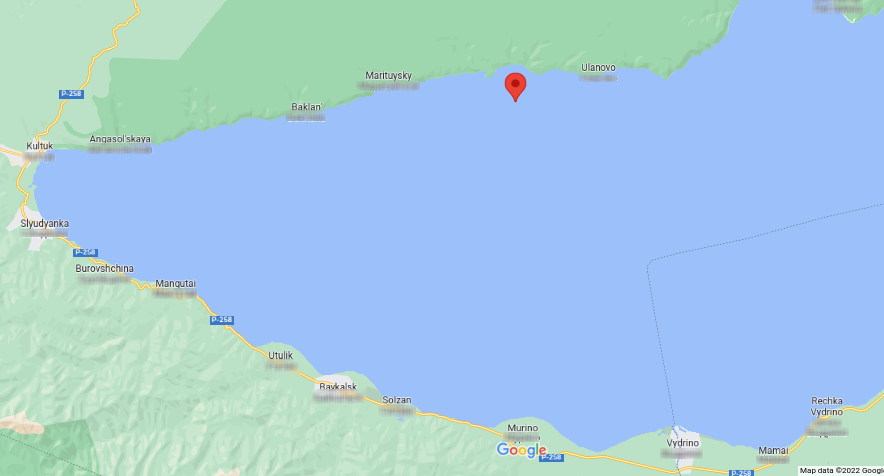}
		\caption{{Google Map screen shot of the Baikal-GVD location and its position relative to the lake shore in the north.}}
		\label{fig:googlemap}
	\end{figure}

	The Baikal-GVD experiment \cite{2021Baikal} is located at Lake Baikal (as shown in Figure \ref{fig:googlemap}), Russia very near to the location of the NT-200 experiment was there. The site is located approximately 4 km away from the shore where the depth is nearly constant at \mbox{1366--1367 m}.  The telescope consists of independent structural units---clusters. After the winter expedition in February--April 2021, the detector includes 8 clusters with the sum effective volume of $\sim$0.4 km$^3$. Each cluster of the telescope consists of 8 strings each carrying 36 optical modules (OM) and calibration systems. OMs are located at depths from 750 to 1275 m with a vertical step of 15 m. The cluster radius is 60 m and the horizontal distance between central strings of neighboring clusters is $\sim$300 m. Baikal-GVD optical module is a glass sphere that hosts the photomultiplier tube (PMT) Hamamatsu R7081-100 with a 10-inch hemispherical photocathode oriented towards the lake floor along with various sensors, readout, high-voltage control electronics, and calibration LEDs.

	The high-energy atmospheric muons produced in the cosmic ray showers emit \linebreak Cherenkov light along the length of its track in the detector which is detected by the OMs. This provides an excellent opportunity to study the atmospheric muon flux underwater at very high energies. The CORSIKA software \cite{Engel2018akg} used for the simulation of cosmic ray showers, provides information on particles produced in the interactions but only at the surface. In the present work, a simulation program based on GEANT4 \cite{geant4ref1} is developed for the propagation of muons inside the water. The northern shore is also constructed in the geometry to account for the deficit in the number of muons arriving from that direction. A similar approach was used by Panchal et al. \cite{2019Neha, 2021NehaDAE} to understand the atmospheric muon background for shallow depth experiments. The aim of this work is to provide a simulation framework for the prediction of underwater fluxes after taking into account the real location of the detector.
	
	\section{Description of Simulation}
	
\begin{figure}[H]
		\includegraphics[width=0.99\textwidth,frame]{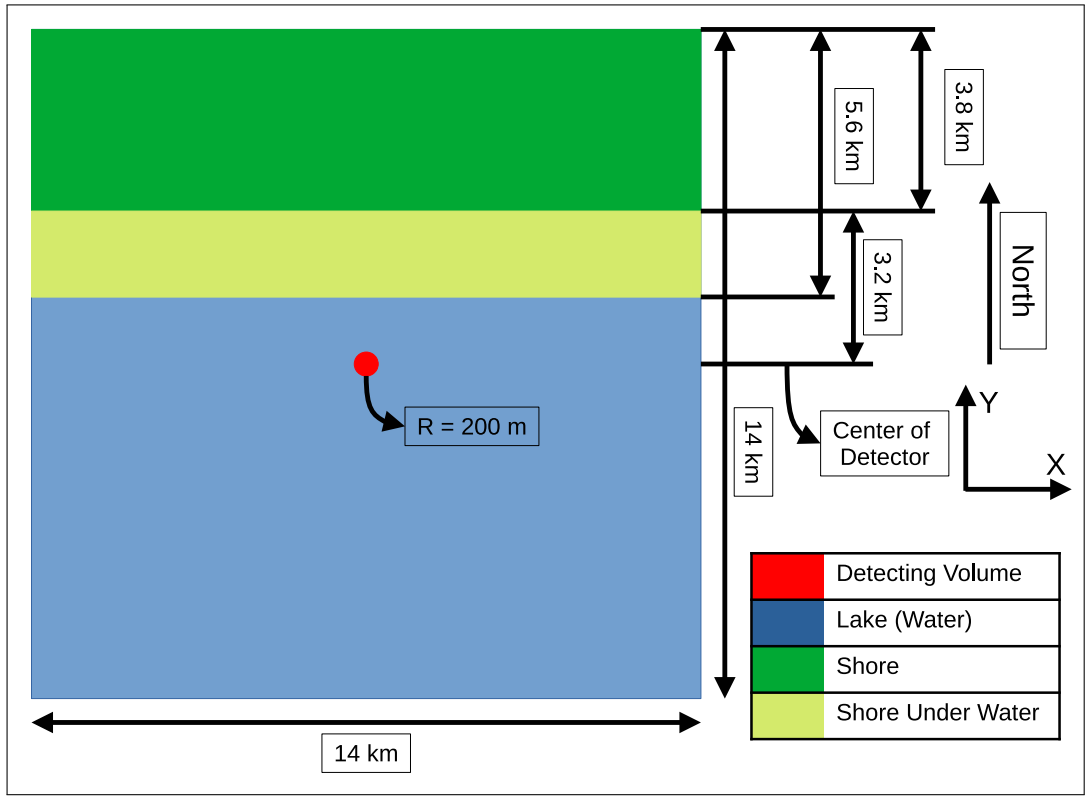}
		\caption{XY view of the schematics of the geometry used in the simulation.}
		\label{fig:shoreSchema1}
	\end{figure}

\begin{figure}[H]
		\includegraphics[width=0.99\textwidth,frame]{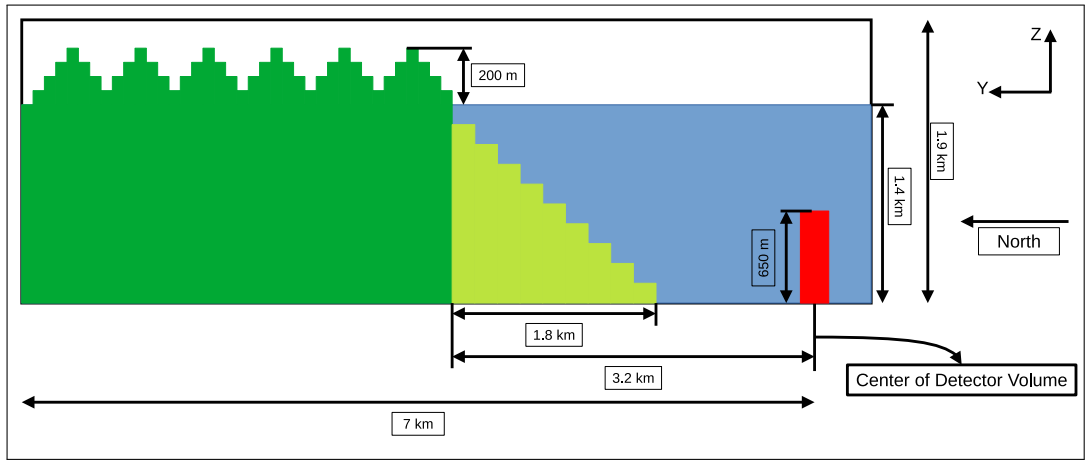}
		\caption{YZ view of the schematics of the geometry used in the simulation.}
		\label{fig:shoreSchema2}
	\end{figure}

	A GEANT4-based simulation approach has been developed to understand the Shore Shadow effect for the Baikal-GVD experiment. The idea is to propagate muons using the GEANT4 framework until they reach the active volume of the Baikal detector from all azimuthal directions. In this simulation, the entire active volume of the  Baikal-GVD detector is considered here as a cylinder\endnote{In this article, this will be referred to as \textbf{detector volume}.} with a radius of 200 m, and height 650 m as shown in Figures \ref{fig:shoreSchema1} and \ref{fig:shoreSchema2}. This ``detector volume'' is surrounded on all sides by water. As the Baikal-GVD detector strings are located 750 m below the surface of the lake, the same has been taken into account in this simulation. The +X- and +Y-axis are considered as the \textit{{east}} and \textit{{north}} direction respective. A flat shoreline has been modeled along the northern boundary at a distance of 3.2 km. The shore has been extrapolated for a further distance of 3.8 km towards the north. The underwater shore is extended for a distance of 1.8 km in steps of 200 m width to approximate a slope of 40$^\circ$. Some of the hills observed on the shore are simulated with an approximation of steps of 100 m  with an increase or decrease in height of 50 m (Figure \ref{fig:shoreSchema2}). For the other three directions, for a distance of approximately 7 km from the detector volume, only water is considered.
	
	\unskip
	
	All the interactions experienced by a muon while traversing through matter have been considered in the \textit{{PhysicsLists}} of GEANT4 to account for the energy loss in the simulation. There are four basic processes of muon interaction that determine muon energy loss and generation of secondary showers in the matter: ionization (including production of high-energy knock-on electrons, or $\delta$-rays), direct production of electron-positron pairs, bremsstrahlung, and inelastic muon interaction with nuclei. At moderate muon energy (E$\sim$100 GeV) practically in the whole region, the main interaction process is the production of knock-on electrons. For high-energy muons, the role of pair production, bremsstrahlung, and inelastic interaction is much more dominant and the energy loss process is essentially stochastic.

	In GEANT4, for the description of muon bremsstrahlung, the results of \cite{Kelner1995} are used that represent improved parameterization of Petrukhin-Shestakov formula \cite{Petrukhin1968} valid for any degree of screening and accounting for the nuclear form factor. Formulae for muon bremsstrahlung on atomic electrons are taken from \cite{Kelner1997}. Bremsstrahlung from target electrons (together with the respective radiative correction to $\mu$-e scattering) is taken into account as a correction term \cite{Kelner1997} to Bhabha formula for knock-on electron production. Electron pair production on the screened nucleus is described by formulae from \cite{Kokoulin1969,Kokoulin1971}. The results from \cite{Kelner1998} are used for the contribution of pair production on electrons. For the cross-section of inelastic muon interaction with nuclei, expression from \cite{Borog1975} is applied.
	
\begin{figure}[H]
		\includegraphics[width=0.6\textwidth]{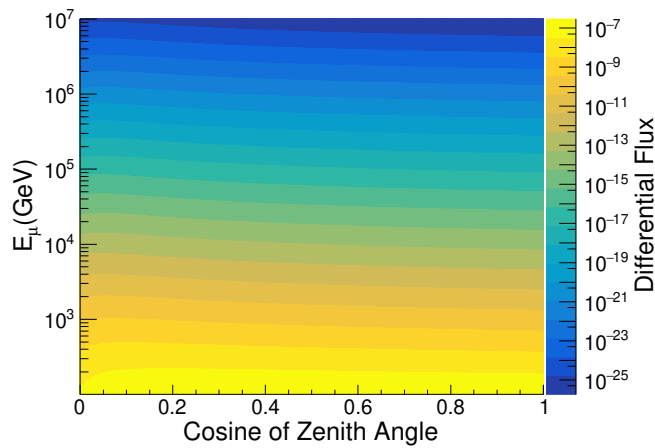}
		\caption{The differential muon flux as a function of cosine of the zenith angle and the energy of the muon as calculated using AMBala at the earth's surface.}
		\label{fig:AMBalaFlux}
	\end{figure}

	The muons are simulated in the energy range of $10^3$--$10^7$ GeV using the cosmic muon flux calculated using the AMBala package \cite{MaleckiP_Pvt}. AMBaLa (Atmospheric Muons in the Baikal Lake) is a package that represents a little collection consisting of 150+ sets of experimental data and theoretical predictions for atmospheric muon fluxes both at the sea level and deep underwater/ice, at typical depths where underwater/ice neutrino telescopes are deployed. In particular, it contains references also data on surface spectra derived from underground experiments. For the present work, the 2019 calculations for double differential atmospheric muon fluxes at the sea level from Sinegovsky et al. \cite{Sinegovsky2019} which uses primary cosmic ray spectrum and composition by Gaisser and Hillas \cite{Gaisser2012} and hadronic cross-section models by Kimel and Mokhov \cite{Kimel1974}. The output of AMBala is the differential muon flux as a function of muon zenith angle and muon energy which is shown in Figure \ref{fig:AMBalaFlux}.

	As the differential flux obtained from the AMBala does not have the information on the azimuthal dependence of the muon flux, for the purpose of the present work, it has been assumed to be uniform. The muon vertices are simulated at the boundaries of the outer box and only muons which have a chance to reach the detector volume, based on its direction and with the assumption that they travel along a straight path are allowed to be propagated. This simulation can be easily modified to use the muon events on the surface generated by CORSIKA as an input to obtain the modified underwater flux.
	
	When the GEANT4 propagates the muon, if the muon reaches the \textit{{detector volume}}, the propagation of the muon is stopped and it is considered as detected for the purpose of this work. At that point, the kinematic variables of the muon namely its Energy (E) and the direction vector of its momentum are recorded. This approach allows computing the modulated muon flux under the surface of the water along with the effect of the shore. This modulated muon flux can be used as input to the full detector simulation and reconstruction chain to compare the data and MC.
	
	\section{Results}
	
\begin{figure}[H]
		\includegraphics[width=0.45\textwidth]{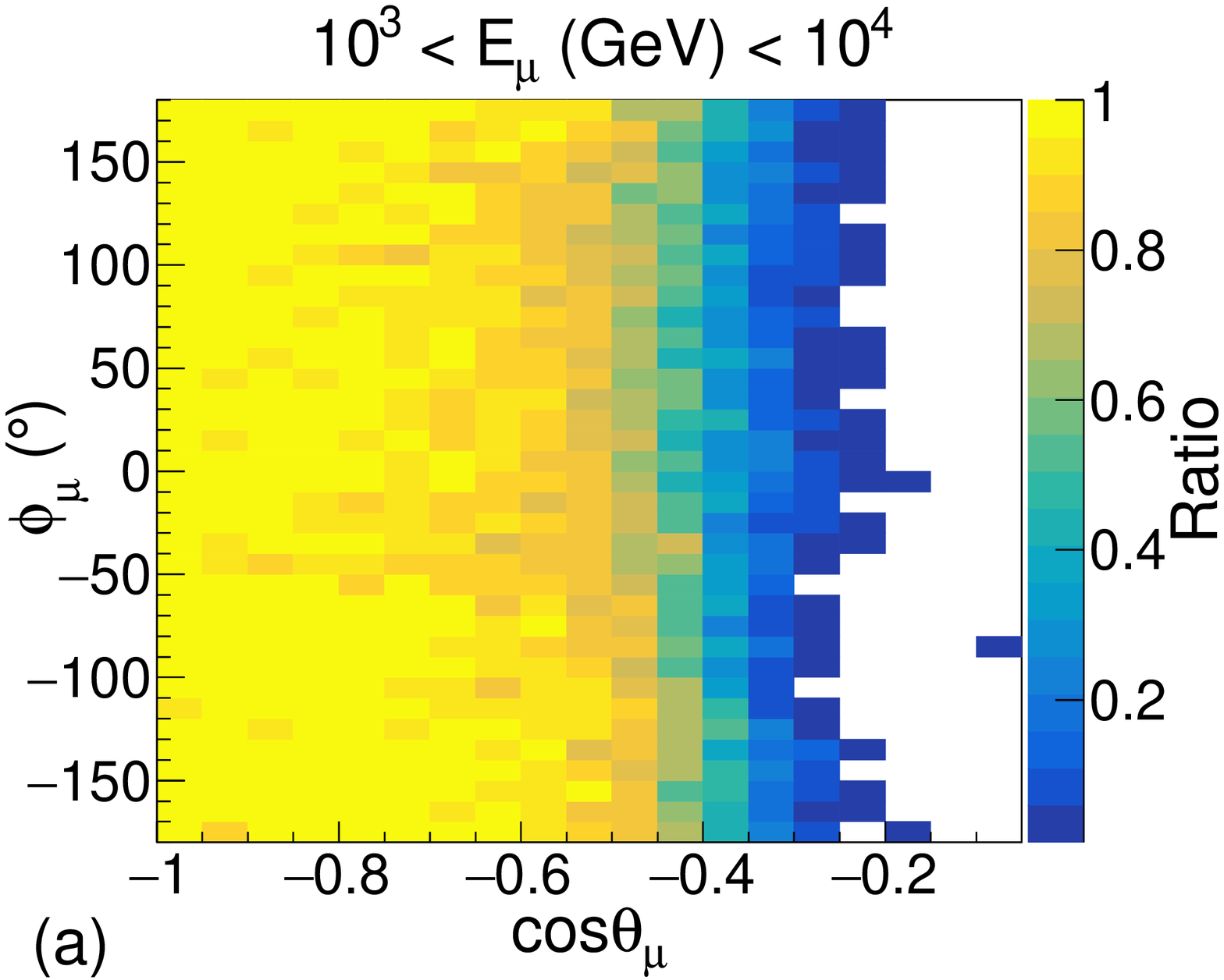}
		\includegraphics[width=0.45\textwidth]{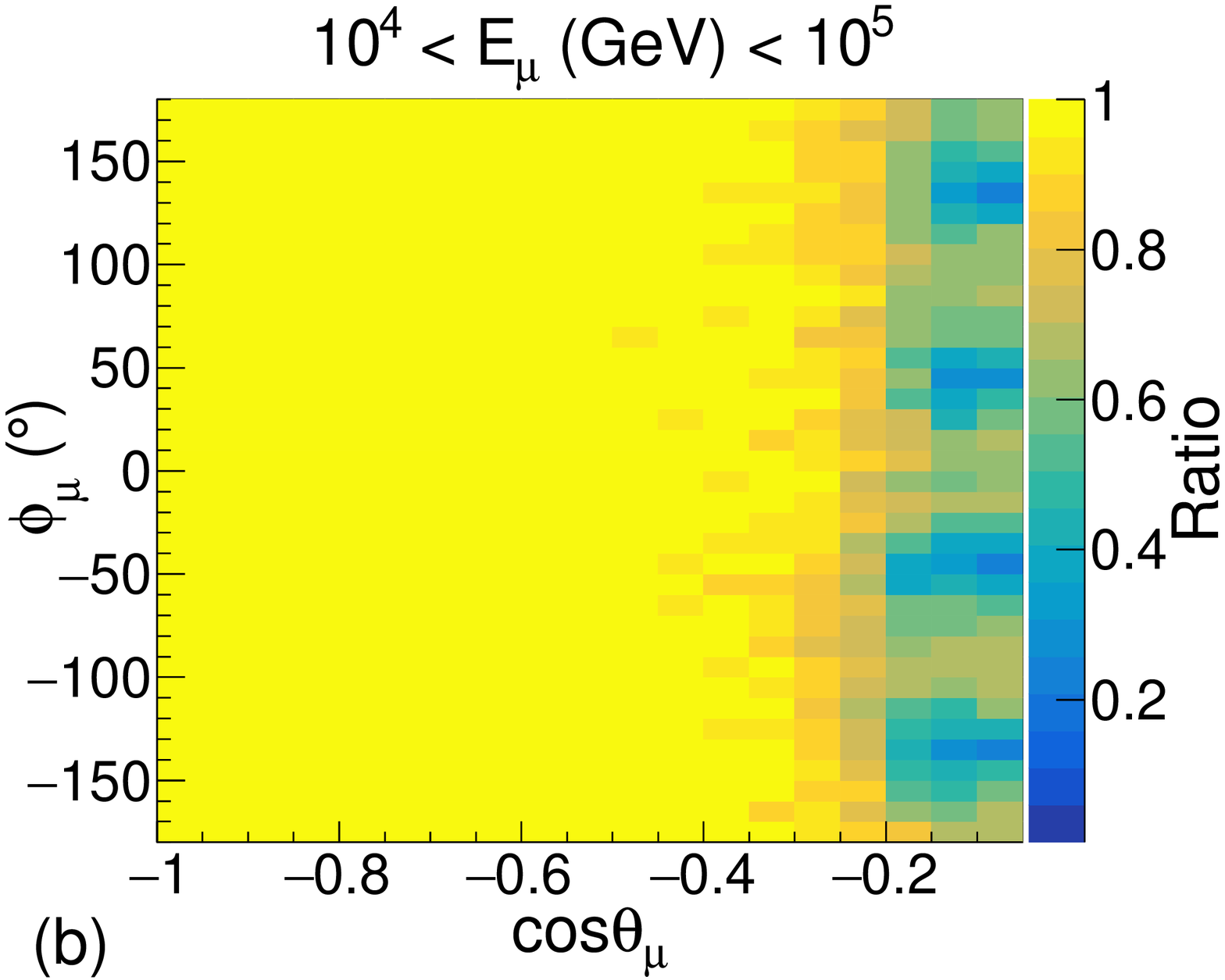}\\
		\includegraphics[width=0.45\textwidth]{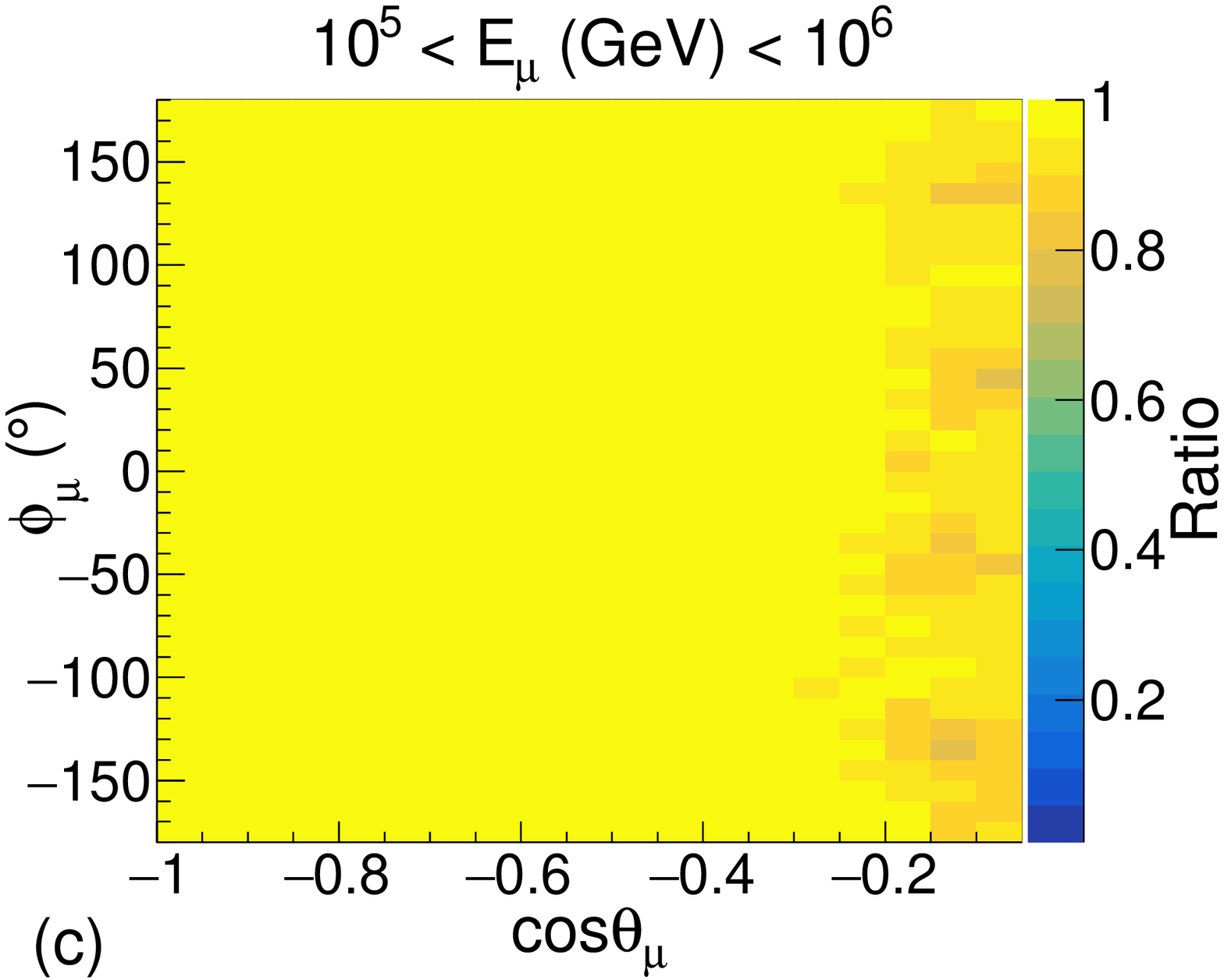}
		\includegraphics[width=0.45\textwidth]{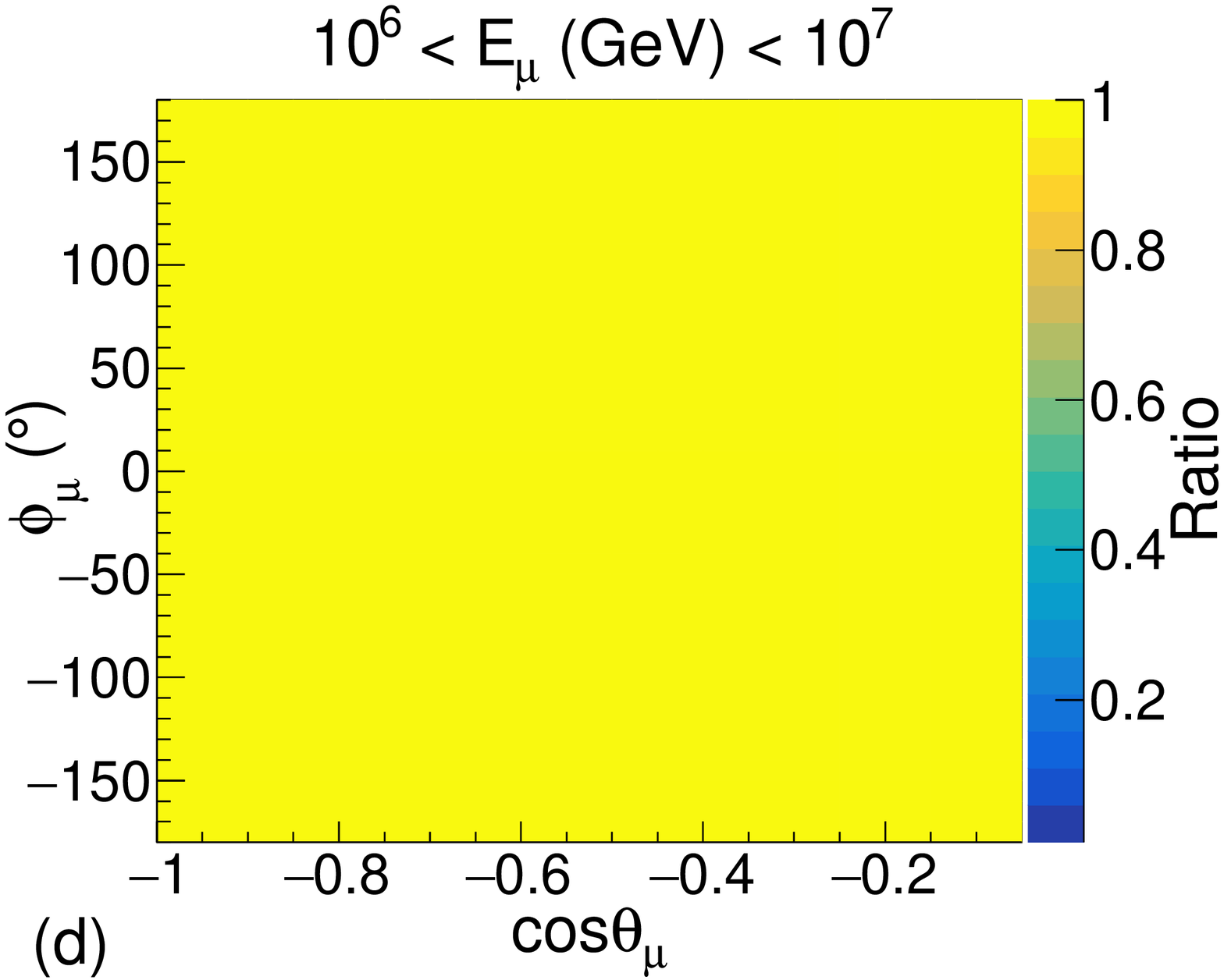}
		\caption{{The ratio of the number of muons detected to the number of muons generated in the simulation without the shore with respect to the muon arrival direction for energy ranges: (a) $10^3$--$10^4$\,GeV, (b) $10^4$--$10^5$\,GeV, (c) $10^5$--$10^6$\,GeV, and (d) $10^6$--$10^7$\,GeV.}}
		\label{fig:resultWithoutShore}
	\end{figure}

\begin{figure}[H]
		\includegraphics[width=0.45\textwidth]{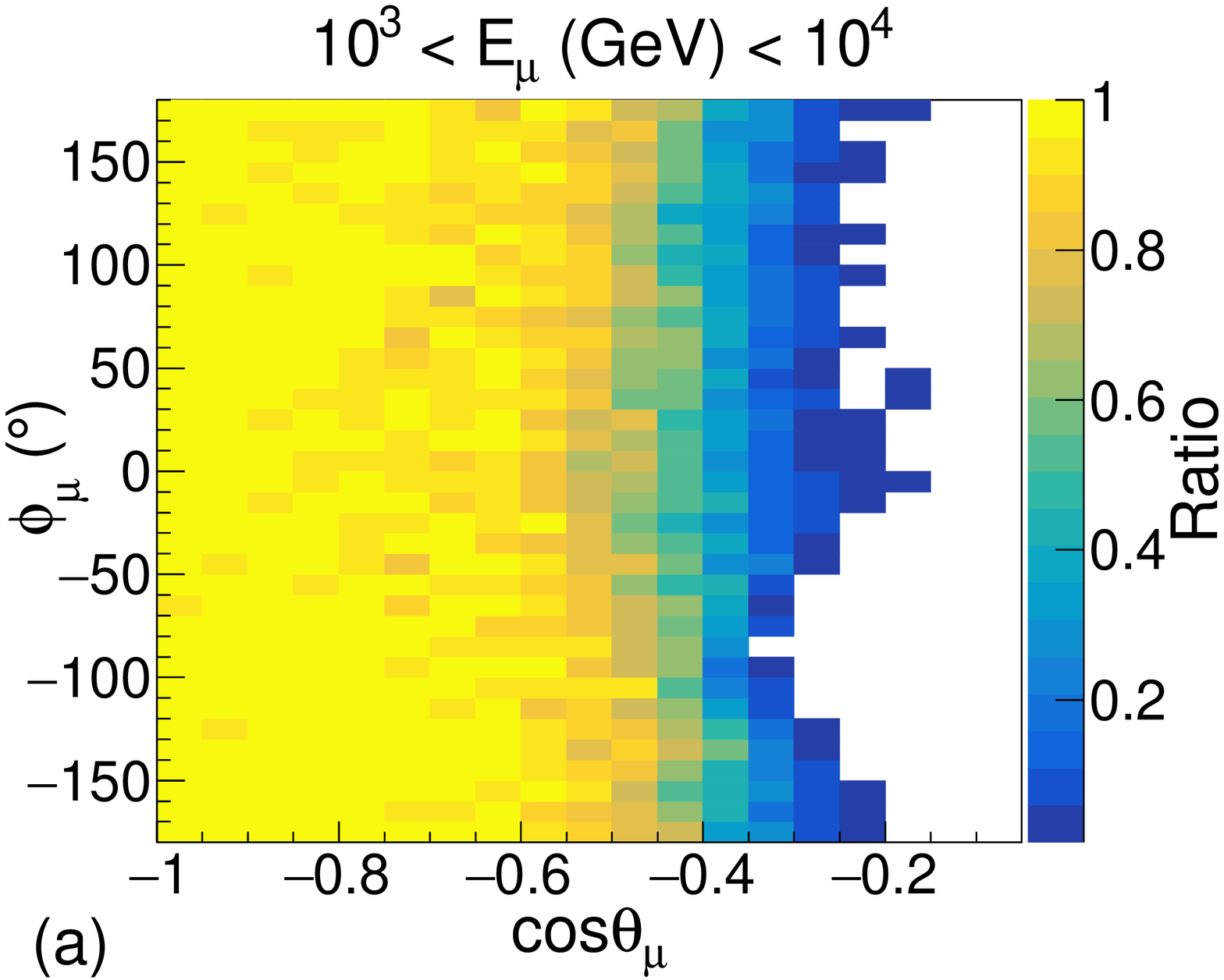}
		\includegraphics[width=0.45\textwidth]{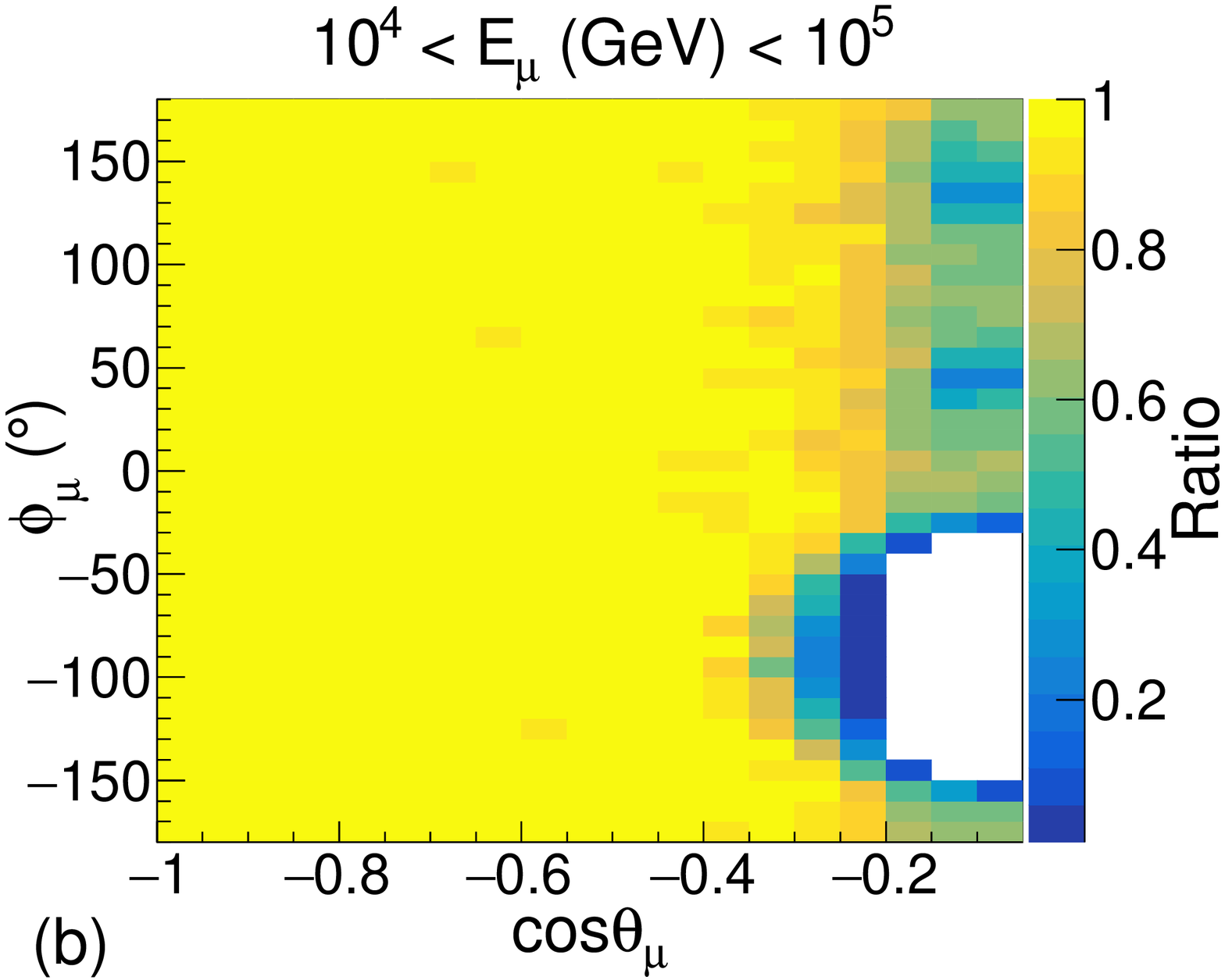}\\
		\includegraphics[width=0.45\textwidth]{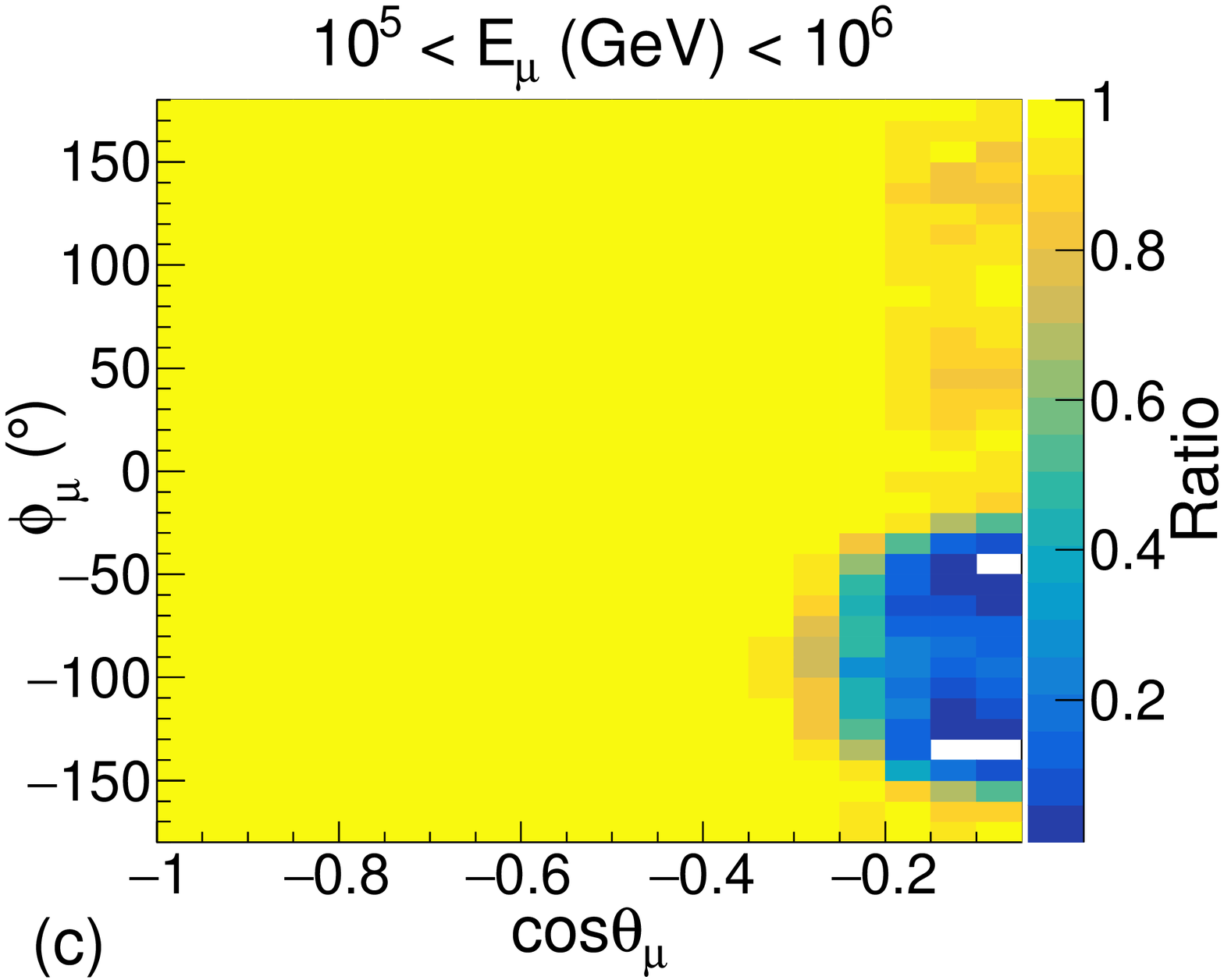}
		\includegraphics[width=0.45\textwidth]{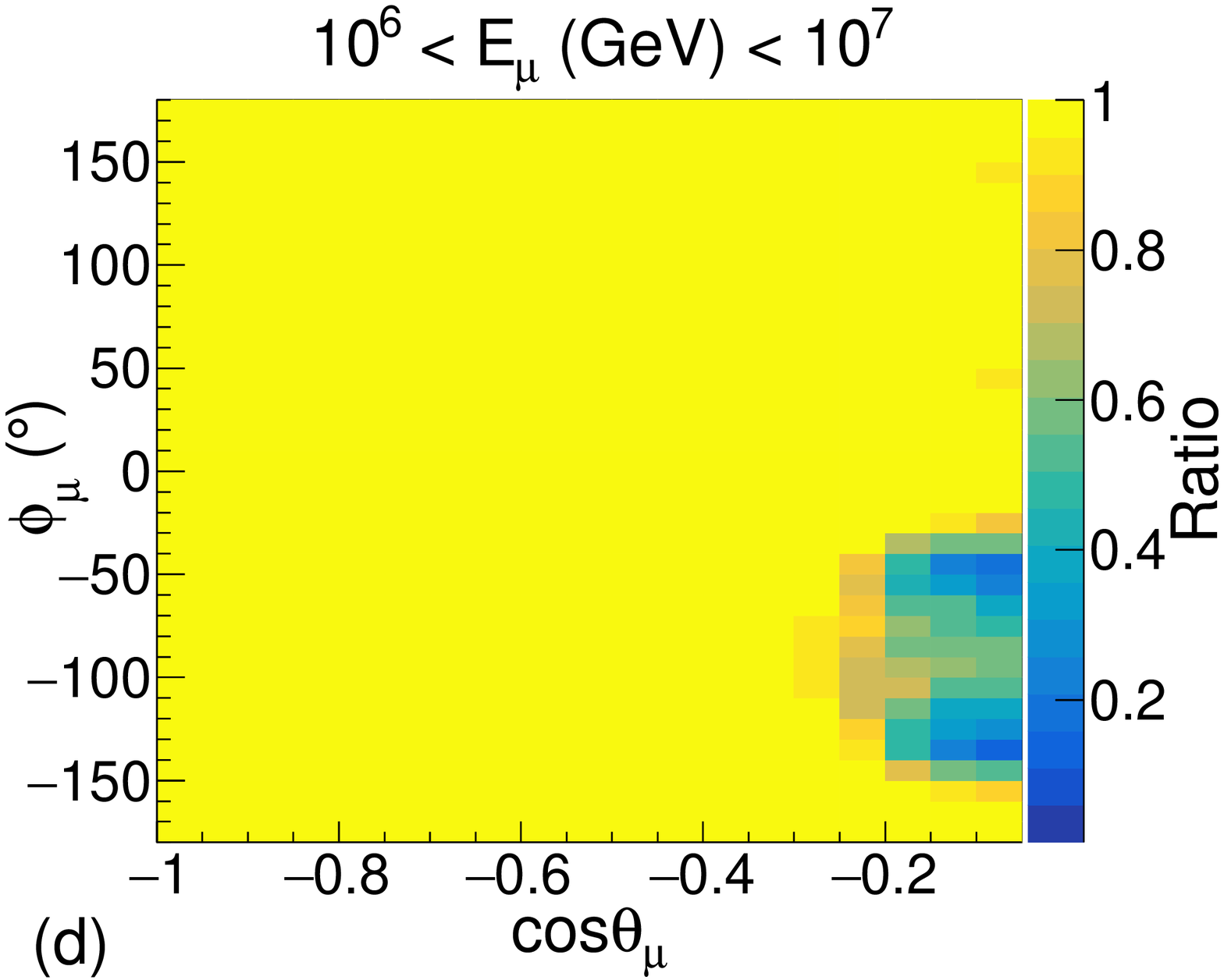}
		\caption{{The ratio of the number of muons detected} to the number of muons generated in the simulation with the shore with respect to the muon arrival direction for energy ranges: (a) $10^3$--$10^4$\,GeV, (b) $10^4$--$10^5$\,GeV, (c) $10^5$--$10^6$\,GeV, and (d) $10^6$--$10^7$\,GeV.}
		\label{fig:resultWithShore}
	\end{figure}

	The simulation is performed once without putting the shore in the geometry and a second time after introducing the shore. The results are presented in Figures \ref{fig:resultWithoutShore} and \ref{fig:resultWithShore}, respectively. In these plots, the ratio of the number of muons detected to the number of muons generated is plotted on the Z-axis to understand its dependence on the direction of the muon. The X- and Y-axes in the plots represent the direction of muon momentum $\theta_\mu$ and $\phi_\mu$ in spherical coordinate systems which are related to the zenith and azimuthal angles of the direction of muon arrival.
	
	\unskip
	
	It can be observed in Figure \ref{fig:resultWithoutShore}a that for muons in the energy range of $10^3$--$10^4$ GeV, significantly fewer muons will be able to reach the detector for cos\,$\theta_\mu>-0.4$ due to the energy loss in the water in the lake. An identical result can be observed in the simulation with the shore in Figure \ref{fig:resultWithShore}a. Hence, for the comparison of the simulation results with the data from Balkanov et al. \cite{2000BaikalNT}, the results of simulation for muon energies above $10^4$ GeV are considered. In Figure \ref{fig:resultWithoutShore}b,c, a reduction in the arrival of a number of muons for cos $\theta_\mu>-0.3$ and for values of $\phi_\mu$ around $-$135$^\circ$, $-$45$^\circ$, 45$^\circ$ and 135$^\circ$ is mainly due to the cuboid-shaped geometry used in the simulation. These four values of $\phi_\mu$ represent the four corners of the square (as looked at the geometry from the top view).
	
\begin{figure}[H]
		\includegraphics[width=0.45\textwidth]{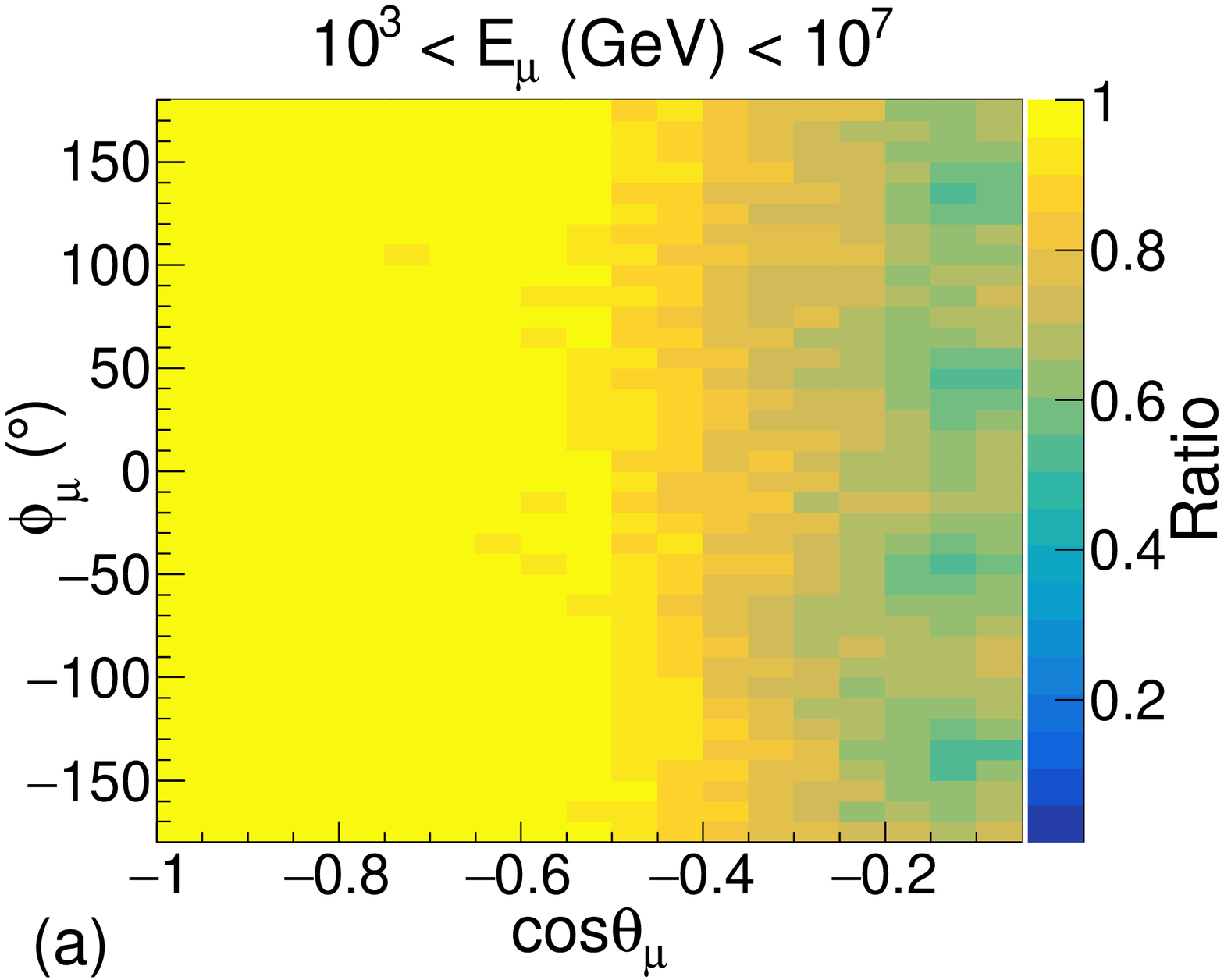}
		\includegraphics[width=0.45\textwidth]{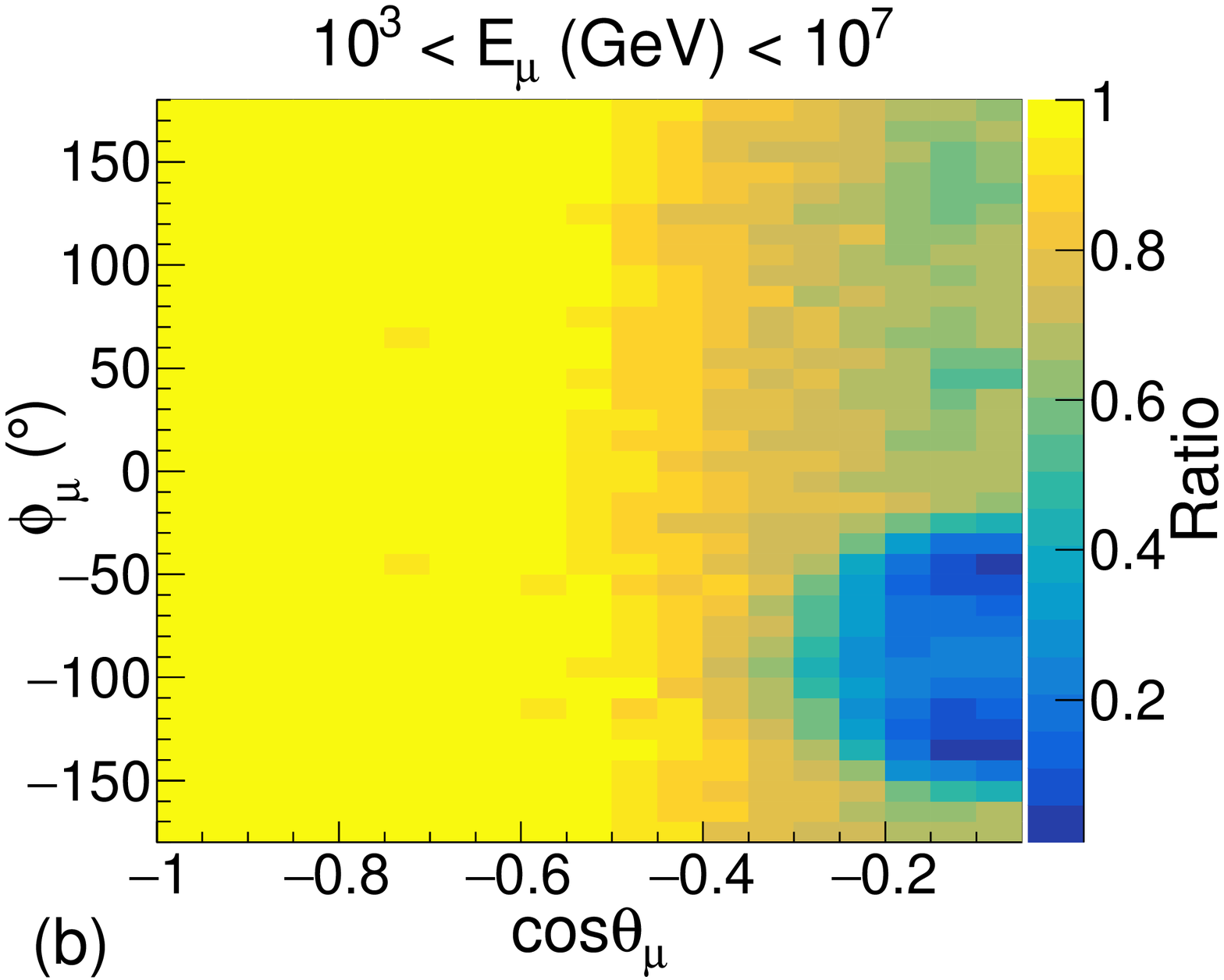}
		\caption{The ratio of the number of muons detected to the number of muons generated in the simulation with the entire energy range with respect to the muon arrival direction. (\textbf{a}) Without Shore (\textbf{b}) With Shore.}
		\label{fig:resultAllCombined}
	\end{figure}

	It can be observed in Figure \ref{fig:resultWithShore}b,c, for  cos $\theta_\mu > -0.35$ and $\phi_\mu\in\left(45^\circ,135^\circ\right)$, there is a signification reduction in the number of muons reaching the detection volume. This clearly represents the effect of the shore in the simulation. For even higher energies of muons, the effect of the shore can still be observed but it is not very prominent as can be seen in Figure \ref{fig:resultWithShore}d. The results of the simulation with the entire energy range of muons are presented in Figure \ref{fig:resultAllCombined}. A significant reduction in the number of muons reaching the detection volume from the direction of the shore can be observed when the shore is introduced in the simulation. The deficit in the number of muons observed for cos $\theta_\mu > 0.4$ in Figure \ref{fig:resultAllCombined}a is due to the muons in the low energy range, i.e., 10$^3$--10$^4$ GeV.
	
\begin{figure}[H]
		\includegraphics[width=0.45\textwidth]{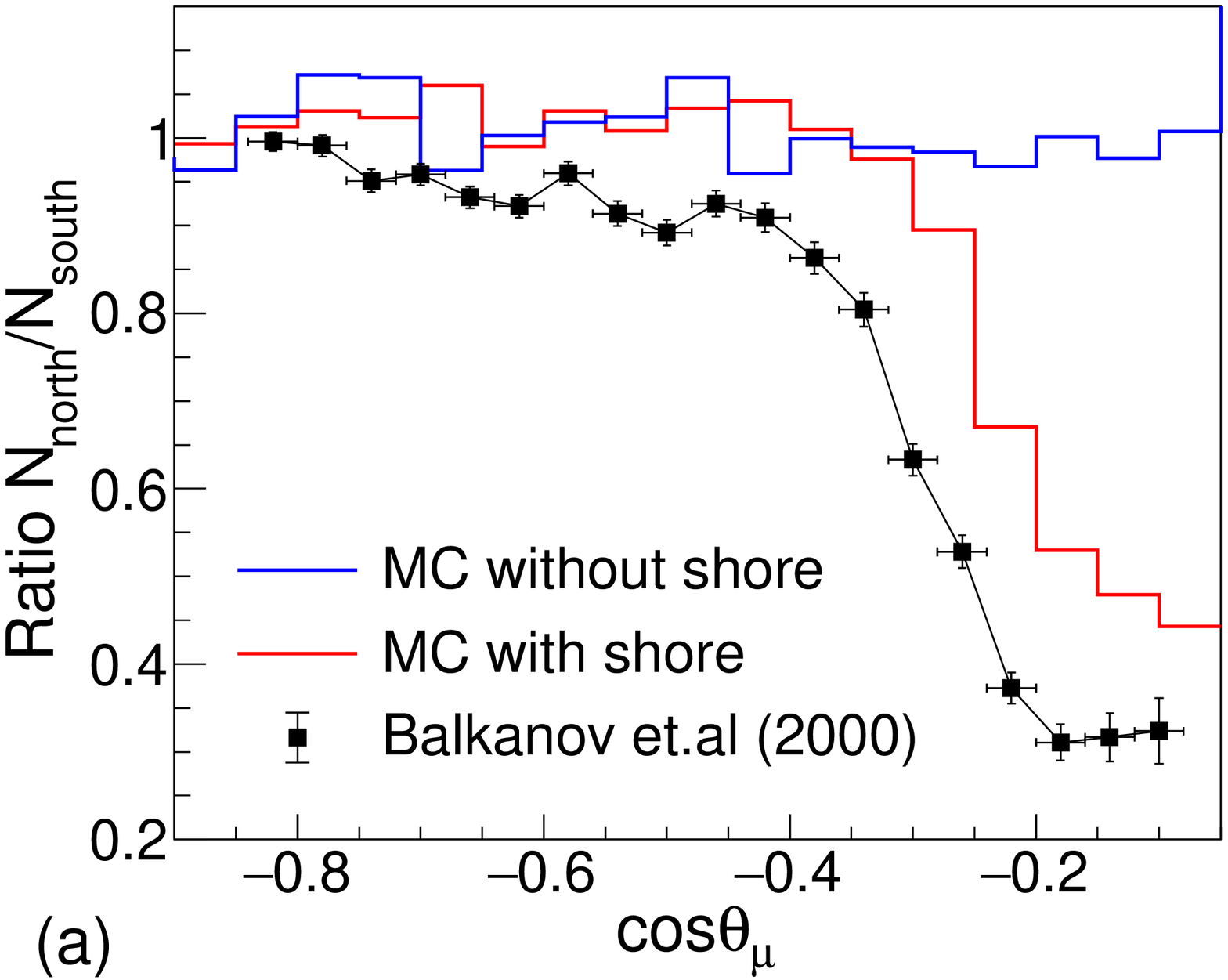}
		\includegraphics[width=0.45\textwidth]{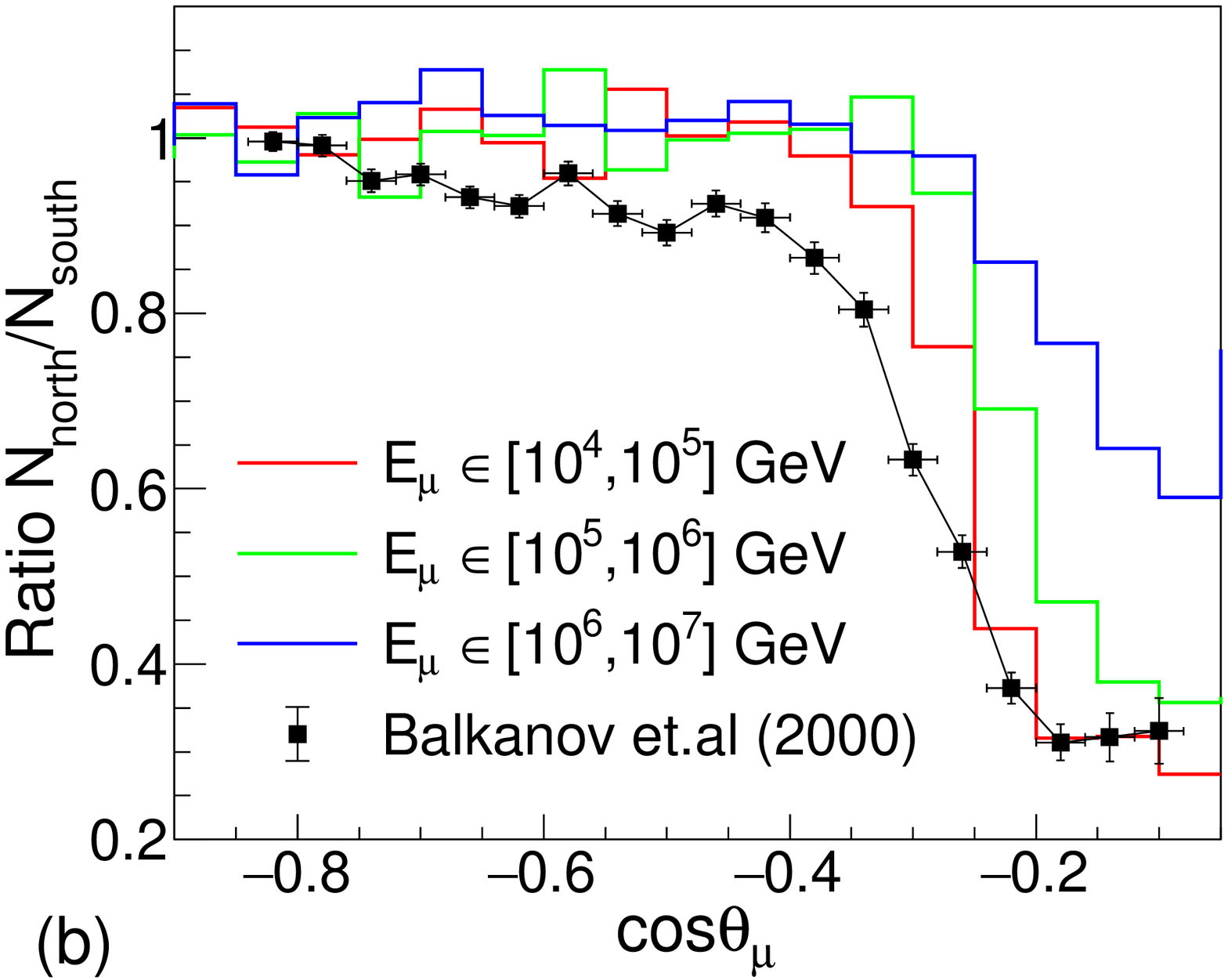}
		\caption{{Comparison of simulation with data presented in} {Balkanov et al.} \cite{2000BaikalNT}. (\textbf{a}) The simulation results with the entire energy range with and without the shore, and (\textbf{b}) the simulation results in different ranges of $E_\mu$.}
		\label{fig:resultDataMC}
	\end{figure}

	In Figure \ref{fig:resultDataMC}, the results of the simulation discussed in this work are compared to the experimental observations presented in Balkanov et al. \cite{2000BaikalNT}. In the plot, $\text{N}_{\text{north}}$ represents the number of muons arriving at the detector from the northside ($\phi_\mu<0$), and $\text{N}_{\text{south}}$ represents the number of muons arriving at the detector from the southside ($\phi_\mu>0$). The ratio of these numbers $\text{N}_{\text{north}}/\text{N}_{\text{south}}$ is plotted as function of cos $\theta_\mu$ to compare the simulation results with data from Balkanov et al. \cite{2000BaikalNT}. As observed in Figure \ref{fig:resultDataMC}a, there is a clear reduction in the number of muons arriving from the north when the two simulations (with and without shore) are compared from cos $\theta_\mu>-0.4$. This observation is consistent with the experimental results of Balkanov et al. \cite{2000BaikalNT}.
	
	\unskip
	
	Although, compared to the data, it is observed that in simulation, marginally more muons are detected from the direction of the shore. The difference in the present simulation and the results of Balkanov et al. \cite{2000BaikalNT} are mainly due to the limitations in the simulation. For simplicity, a straight shoreline is considered in the geometry of the simulation, however, the exact shoreline is not straight. The terrain above the water level in the lake shows some mountains. The terrain above the water level is not considered with full specifications in the simulation. As well as, the shore is simulated with an incline of $45^\circ$ below the water level, but the actual terrain may be less steep and may extend for more than the distance used in the simulation. The default SiO$_2$ material available in the database of the GEANT4 is used as the material to simulate the shore. The density and the chemical composition of the soil on the shore of the Baikal Lake will be certainly different from the approximation used here. For high-energy muons, $\left(\text{E}_\mu>10^4\text{ GeV}\right)$, may give even more attenuation in the number of muons arriving from the north direction which is observed in the data. In Figure \ref{fig:resultDataMC}b, it can be observed that the ratio of events for muons in the energy range of $10^4$--$10^5$ GeV follows more closely with data as compared to the muons of higher energies. This higher energy contribution to the full energy range may also contribute to this little discrepancy between the data and the simulation. At these high energies of the muon, the energy loss is dominated by the radiative processes, which may also contribute to the uncertainties in the simulation.
	
	\section{Summary}
	
	A simulation program has been developed to propagate muons inside the water. It can be used to predict the underwater atmospheric muon flux for the experiment by using the known muon flux on the surface. This approach can also be used to propagate atmospheric muons from generators like CORSIKA which provide the kinematics of muons at the earth's surface through long distances in matter like water, earth, ice, etc for neutrino telescope-like experiments to study very high-energy atmospheric muons. In this work, the approach was used to study the shore shadow effect in the Baikal Lake experiments. The result of the simulation is consistent with the shadow effect observed in the measurement of underwater muon flux in the NT-96 experiment \cite{2000BaikalNT}.
	
	\vspace{6pt} 
	
	
	\authorcontributions{Conceptualization, A.B. and P.M.; methodology, software, validation and formal analysis, A.B.; writing---original draft preparation, A.B.; writing---review and editing, D.G. and P.M.; supervision, D.G. All authors have read and agreed to the published version of the manuscript.}
	
	\funding{This research received no external funding.}
	
	\institutionalreview{Not applicable.}
	
	\informedconsent{Not applicable.}
	
	\dataavailability{Not applicable.} 
	
	\acknowledgments{The research presented in this paper was partially supported by \textbf{{plgbgvd2021}}, \textbf{{plgbgvd2022}} grants.}
	
	\conflictsofinterest{The authors declare no conflict of interest. The funders had no role in the design of the study; in the collection, analyses, or interpretation of data; in the writing of the manuscript, or in the decision to publish the results.} 
	
	\abbreviations{Abbreviations}{
		The following abbreviations are used in this manuscript:\\
		
		\noindent 
		\begin{tabular}{@{}ll}
			MeV & Mega Electron-Volt ($10^6$ Electron-Volt)\\
			GeV & Giga Electron-Volt ($10^9$ Electron-Volt)\\
			TeV & Tera Electron-Volt ($10^{12}$ Electron-Volt)\\
			NT & Neutrino Telescope\\
			OM & Optical Modules\\
			PMT & Photo Multiplier Tube\\
			GVD & Gigaton Volume Detector\\
			EAS & Extended Air Showers\\
			LED & Light Emitting Diode\\
			Baikal-GVD & Baikal-Gigaton Volume Detector\\
			AMBala & Atmospheric Muons in the Baikal Lake\\
			MC & Monte Carlo
		\end{tabular}
	}

	\begin{adjustwidth}{-\extralength}{0cm}
		\printendnotes[custom]
		\reftitle{References}
		
		

	\end{adjustwidth}

	%
	
	

\begin{thebibliography}{999}
			
			\bibitem[Balkanov \em{et al.}(2000)Balkanov, Belolaptikov, Bezrukov, Budnev,
			Chensky, Danilchenko, Dzhilkibaev, Domogatsky, Doroshenko, Fialkovsky, and
			et al.]{2000BaikalNT}
			Balkanov, V.A.; Belolaptikov, I.A.; Bezrukov, L.B.; Budnev, N.M.; Chensky,
			A.G.; Danilchenko, I.A.; Dzhilkibaev, Z.A.M.; Domogatsky, G.V.; Doroshenko,
			A.A.; Fialkovsky, S.V.;  et al.
			\newblock Lake Baikal neutrino experiment: Selected results.
			\newblock {\em Phys. At. Nucl.} {\bf 2000}, {\em 63}, 951--961. https://doi.org/10.1134/1.855731.
			
			\bibitem[Gaisser(1990)]{GaisserBook}
			Gaisser, T.K.
			\newblock {\em {Cosmic Rays and Particle Physics}}; Cambridge University
			Press: Cambridge, UK, 1990.
			
			\bibitem[Mondal \em{et al.}(2020)Mondal, Datar, Majumder, Mondal, Pethuraj,
			Ravindran, and Satyanarayana]{2020Surya}
			Mondal, S.; Datar, V.M.; Majumder, G.; Mondal, N.K.; Pethuraj, S.; Ravindran,
			K.C.; Satyanarayana, B.
			\newblock Study of particle multiplicity of cosmic ray events using 2 m × 2 m
			resistive plate chamber stack at IICHEP-Madurai.
			\newblock {\em Exp. Astron.} {\bf 2020}, {\em 51}, 17--32. https://doi.org/10.1007/s10686-020-09685-6.
			
			\bibitem[Pethuraj \em{et al.}(2020)Pethuraj, Majumder, Datar, Mondal,
			Ravindran, and Satyanarayana]{2020Pethuraj}
			Pethuraj, S.; Majumder, G.; Datar, V.M.; Mondal, N.K.; Ravindran, K.C.;
			Satyanarayana, B.
			\newblock Measurement of azimuthal dependent muon flux by 2 m × 2 m RPC stack
			at IICHEP-Madurai.
			\newblock {\em Exp. Astron.} {\bf 2020}, {\em 49}, 141–157. https://doi.org/10.1007/s10686-020-09655-y.
			
			\bibitem[Safronov \em{et al.}(2021)Safronov, Allakhverdyan, Avrorin, Avrorin,
			Aynutdinov, Bannasch, Bardáčová, Belolaptikov, Brudanin, Budnev, and
			et al.]{2021Baikal}
			Safronov, G.; Allakhverdyan, V.; Avrorin, A.; Avrorin, A.; Aynutdinov, V.;
			Bannasch, R.; Bardáčová, Z.; Belolaptikov, I.; Brudanin, V.; Budnev, N.;
			et al.
			\newblock Performance of the muon track reconstruction with the Baikal-GVD
			neutrino telescope. In Proceedings of 37th International Cosmic Ray Conference---PoS(ICRC2021), {Online, 12--23 July 2021}. https://doi.org/10.22323/1.395.1080.
			
			\bibitem[Engel \em{et al.}(2019)Engel, Heck, Huege, Pierog, Reininghaus, Riehn,
			Ulrich, Unger, and Veberi\v{c}]{Engel2018akg}
			Engel, R.; Heck, D.; Huege, T.; Pierog, T.; Reininghaus, M.; Riehn, F.; Ulrich,
			R.; Unger, M.; Veberi\v{c}, D.
			\newblock {Towards a Next Generation of CORSIKA: A Framework for the Simulation
				of Particle Cascades in Astroparticle Physics}.
			\newblock {\em Comput. Softw. Big Sci.} {\bf 2019}, {\em 3}, 2. https://doi.org/10.1007/s41781-018-0013-0.
			
			\bibitem[Agostinelli \em{et al.}(2003)Agostinelli, Allison, Amako, Apostolakis,
			Araujo, Arce, Asai, Axen, Banerjee, Barrand, Behner, Bellagamba, Boudreau,
			Broglia, Brunengo, Burkhardt, Chauvie, Chuma, Chytracek, Cooperman, Cosmo,
			Degtyarenko, Dell'Acqua, Depaola, Dietrich, Enami, Feliciello, Ferguson,
			Fesefeldt, Folger, Foppiano, Forti, Garelli, Giani, Giannitrapani, Gibin,
			Cadenas, González, Abril, Greeniaus, Greiner, Grichine, Grossheim, Guatelli,
			Gumplinger, Hamatsu, Hashimoto, Hasui, Heikkinen, Howard, Ivanchenko,
			Johnson, Jones, Kallenbach, Kanaya, Kawabata, Kawabata, Kawaguti, Kelner,
			Kent, Kimura, Kodama, Kokoulin, Kossov, Kurashige, Lamanna, Lampén, Lara,
			Lefebure, Lei, Liendl, Lockman, Longo, Magni, Maire, Medernach, Minamimoto,
			de Freitas, Morita, Murakami, Nagamatu, Nartallo, Nieminen, Nishimura,
			Ohtsubo, Okamura, O'Neale, Oohata, Paech, Perl, Pfeiffer, Pia, Ranjard,
			Rybin, Sadilov, Salvo, Santin, Sasaki, Savvas, Sawada, Scherer, Sei,
			Sirotenko, Smith, Starkov, Stoecker, Sulkimo, Takahata, Tanaka, Tcherniaev,
			Tehrani, Tropeano, Truscott, Uno, Urban, Urban, Verderi, Walkden, Wander,
			Weber, Wellisch, Wenaus, Williams, Wright, Yamada, Yoshida, and
			Zschiesche]{geant4ref1}
			Agostinelli, S.; Allison, J.; Amako, K.; Apostolakis, J.; Araujo, H.; Arce, P.;
			Asai, M.; Axen, D.; Banerjee, S.; Barrand, G.;  et al.
			\newblock Geant4---A simulation toolkit.
			\newblock {\em Nucl. Instruments Methods Phys. Res. Sect. A Accel. Spectrom. Detect. Assoc. Equip.} {\bf 2003},
			{\em 506}, 250--303. https://doi.org/10.1016/S0168-9002(03)01368-8.
			
			\bibitem[Panchal \em{et al.}(2019)Panchal, Majumder, and Datar]{2019Neha}
			Panchal, N.; Majumder, G.; Datar, V.
			\newblock Simulation of muon-induced neutral particle background for a shallow
			depth Iron Calorimeter detector.
			\newblock {\em J. Instrum.} {\bf 2019}, {\em
				14}, P02032. https://doi.org/10.1088/1748-0221/14/02/p02032.
			
			\bibitem[Panchal \em{et al.}(2021)Panchal, Majumder, and Datar]{2021NehaDAE}
			Panchal, N.; Majumder, G.; Datar, V.M.
			\newblock Simulation Studies for a Shallow Depth ICAL.
			In \emph{XXIII DAE High Energy Physics Symposium}; Springer Singapore: Singapore, 2021; pp. 793--799.
			
			\bibitem[Kelner \em{et al.}(1995)Kelner, Kokoulin, and Petrukhin]{Kelner1995}
			Kelner, S.R.; Kokoulin, R.P.; Petrukhin, A.A.
			\newblock \emph{About Cross Section for High-Energy Muon Bremsstrahlung}; 
			\newblock Technical Report; Moscow Engineering Physics Institute; Report No. MEPHI-95-24; {1995}.
			
			\bibitem[Petrukhin and Shestakov(1968)]{Petrukhin1968}
			Petrukhin, A.A.; Shestakov, V.V.
			\newblock The influence of the nuclear and atomic form factors on the muon
			bremsstrahlung cross section.
			\newblock {\em Can. J. Phys.} {\bf 1968}, {\em 46}, S377--S380. https://doi.org/10.1139/p68-251.
			
			\bibitem[Kelner \em{et al.}(1997)Kelner, Kokoulin, and Petrukhin]{Kelner1997}
			Kelner, S.R.; Kokoulin, R.P.; Petrukhin, A.A.
			\newblock {Bremsstrahlung from muons scattered by atomic electrons}.
			\newblock {\em Phys. At. Nucl.} {\bf 1997}, {\em 60}, 576--583. https://doi.org/101134/1854854.
			
			\bibitem[Kokoulin and Petrukhin(1969)]{Kokoulin1969}
			Kokoulin, R.P.; Petrukhin, A.A.
			\newblock Analysis of the cross-section of direct pair production by fast
			muons.
			In {Proceedings of the} 11th International Conference on Cosmic Rays (ICRC11), {Budapest, Hungary, 25 August--4 September 1969}; p. 277.
			
			\bibitem[Kokoulin and Petrukhin(1971)]{Kokoulin1971}
			Kokoulin, R.P.; Petrukhin, A.A.
			\newblock {Influence of the Nuclear Formfactor on the Cross-Section of Electron
				Pair Production by High Energy Muons}.
			In {Proceedings of the} 12th International Cosmic Ray Conference (ICRC12),  {Hobart, Australia, 16--25 August 1971}; Volume 6, p. 2436.
			
			\bibitem[Kelner(1998)]{Kelner1998}
			Kelner, S.R.
			\newblock {Pair production in collisions between muons and atomic electrons}.
			\newblock {\em Phys. At. Nucl.} {\bf 1998}, {\em 61}, 448--456. https://doi.org/101134/1855093.
			
			\bibitem[Borog and Petrukhin(1975)]{Borog1975}
			Borog, V.V.; Petrukhin, A.A.
			\newblock {The cross-section of the nuclear interaction of high energy muons}.
			In {Proceedings of the} 1975 14th International Conference on Cosmic Rays, {Munich, Germany, 15--29 August 1975}; pp. 1949--1954.
			
			\bibitem[Ma{l}ecki()]{MaleckiP_Pvt}
			Ma{l}ecki, P. {Institute of Nuclear Physics Polish Academy of Sciences, Krakow, Poland.}
			\newblock AMBala Package. Private Communication, {2021.}
			
			\bibitem[Kochanov \em{et al.}(2019)Kochanov, Morozova, Sinegovskaya, and
			Sinegovsky]{Sinegovsky2019}
			Kochanov, A.A.; Morozova, A.D.; Sinegovskaya, T.S.; Sinegovsky, S.I.
			\newblock High-energy atmospheric muon flux calculations in comparison with
			recent measurements.
			\newblock {\em J. Phys. Conf. Ser.} {\bf 2019}, {\em
				1181}, 012054. https://doi.org/10.1088/1742-6596/1181/1/012054.
			
			\bibitem[Gaisser(2012)]{Gaisser2012}
			Gaisser, T.K.
			\newblock Spectrum of cosmic-ray nucleons, kaon production, and the atmospheric
			muon charge ratio.
			\newblock {\em Astropart. Phys.} {\bf 2012}, {\em 35}, 801--806. https://doi.org/10.1016/j.astropartphys.2012.02.010.
			
			\bibitem[Kimel and Mokhov(1974)]{Kimel1974}
			Kimel, L.R.; Mokhov, N.V.
			\newblock Particle distributions in the 10$^2$--10$^{12}$ eV energy range
			initiated by high-energy hadrons in dense media.
			\newblock {\em {Izvestiya Vysshikh Uchebnykh Zavedenii, Fizika}} {\bf 1974}, {\em
				10}, 17--23.
			
		\end{thebibliography}
\end{document}